\documentclass[english,preprint,aps,pra,superscriptaddress,showpacs,floatfix]{revtex4-1}
\usepackage{graphics,epsfig,graphicx}
\usepackage{amsbsy}
\usepackage{amsmath}
\usepackage{bm}
\usepackage{amsfonts}
\usepackage{cancel}
\usepackage{multirow}
\usepackage{babel}
\newcommand {\be}{\begin{equation}}
\newcommand {\ee}{\end{equation}}
\newcommand {\bea}{\begin{eqnarray}}
\newcommand {\ea}{\end{eqnarray*}}
\newcommand {\ba}{\begin{eqnarray*}}
\newcommand {\eea}{\end{eqnarray}}

\begin{document}

\title{ Recombination rates from potential models close to the unitary limit} 

\author{E. Garrido}
\affiliation{Instituto de Estructura de la Materia, CSIC, Serrano 123, E-28006
Madrid, Spain}
\author{M. Gattobigio}
\affiliation{Universit\'e de Nice-Sophia Antipolis, Institut Non-Lin\'eaire de
Nice,  CNRS, 1361 route des Lucioles, 06560 Valbonne, France }
\author{A. Kievsky} 
\affiliation{Istituto Nazionale di Fisica Nucleare, Largo Pontecorvo 3, 56100 Pisa, Italy}

\begin{abstract}

We investigate universal behavior in the recombination rate of three bosons
close to threshold. Using the He-He system as a reference, we solve
the three-body Schr\"odinger equation above the dimer threshold 
for different potentials
having large values of the two-body scattering length $a$. To this aim
we use the hyperspherical adiabatic expansion and we extract the $S$-matrix through
the integral relations recently derived. The results are compared
to the universal form,
$\alpha\approx 67.1\sin^2[s_0\ln(\kappa_*a)+\gamma]$, for different
values of $a$ and selected values of the three-body parameter $\kappa_*$. 
A good agreement with the universal formula is obtained after
introducing a particular type of finite-range corrections, which
have been recently proposed by two of the authors in 
A.~Kievsky and M.~Gattobigio, Phys. Rev. A {\bf 87}, 052719 (2013).
Furthermore, we analyze the validity of the
above formula in the description of a very different system:
neutron-neutron-proton recombination. Our analysis confirms the universal character of the process
in systems of very different scales having a large two-body scattering length. 
\end{abstract}

\maketitle

\section{Introduction.} 

The system of three identical bosons having a large two-body scattering length
has been the subject of intense investigations in recent years. As 
shown by  Efimov in a sequence of 
papers~\cite{efimov:1970_phys.lett.b,efimov:1971_sov.j.nucl.phys.},
the three-body spectrum consists of
an infinite number of states that accumulate to zero with the ratio between
two consecutive states being $E_3^{n+1}/E_3^n = \text{e}^{-2\pi/s_0}$.
In other words, in the limit $a\rightarrow\infty$ the two-body system
shows a continuous scale invariance that
 is broken in the $s$-wave three-body sector of a
bosonic system. The residual symmetry is the discrete scale invariance (DSI);
namely, the physics is invariant under the rescaling $r\rightarrow
\Lambda^n r$, where the constant is usually written
$\Lambda=\text{e}^{\pi/s_0}$, with $s_0\approx 1.00624$ an universal number
that characterises a system of three-identical bosons.

The DSI constrains the form of the observables to be log-periodic functions of
the control parameters. One example is
the atom-dimer scattering length which has the general form
\begin{equation}
a_{AD}/a=d_1 +d_2\tan[s_0\ln(a\kappa_*)+d_3]\,,
\label{eq:a_AD}
\end{equation}
where $d_1,d_2,d_3$ are universal constants~\cite{braaten:2006_physicsreports}. 
For atom-dimer collisions below the dimer breakup threshold, DSI
imposes the following universal form for the effective range function
\begin{equation}
ka\cot\delta=c_1(ka) +c_2(ka)\cot[s_0\ln(a\kappa_*)+\phi(ka)]\,,
\label{eq:cotdelta}
\end{equation}
with $\delta$ the atom-dimer phase-shift and $c_1,c_2,\phi$ universal functions
of the dimensionless variable $ka$, where
$k^2=(4/3)E_i/(\hbar^2/m)$, being $E_i$ the incident energy in the atom-dimer
center of mass frame, and $m$ the boson mass.
It is well known that for $k\rightarrow 0$ the effective range function
satisfies the limit $ka\cot\delta\rightarrow -a/a_{AD}$,  which implies that at $k=0$
the constants $d_1,d_2,d_3$ and $c_1(0),c_2(0),\phi(0)$
are related by simple trigonometric relations. A parametrization of the
universal constants and functions can be found in
Ref.~\cite{braaten:2006_physicsreports}. 

The DSI also constrains the
form of the $S$-matrix for collisions above the dimer threshold, leading to the
following peculiar form for the recombination rate at 
threshold~\cite{nielsen:1999_phys.rev.lett.,esry:1999_phys.rev.lett.,petrov:2005_}
\begin{equation}
K_3=\frac{128\pi^2(4\pi-3\sqrt{3})}{\sinh^2(\pi s_0)+\cosh^2(\pi s_0)
       \cot^2[s_0\ln(a\kappa_*)+\gamma]}\frac{\hbar a^4}{m}\,,
\label{eq:recomb}
\end{equation}
that, using the large value of the factor $e^{2\pi s_0}\approx 515$,
can be approximated by
\begin{equation}
  K_3=\alpha\, a^4\hbar/m
  \approx 67.1\sin^2[s_0\ln(\kappa_*a)+\gamma] a^4\hbar/m \,, 
  \label{eq:recombSine}
\end{equation}
where
$\gamma = 1.16 $~\cite{braaten:2006_physicsreports}.

In this work we study in detail the universal behavior
of $\alpha$ by solving the Schr\"odinger equation for a family of
attractive two-body gaussian potentials describing the He-He system.
These potentials are constructed to reproduce the two-body binding energy $E_2$,
the two-body scattering length $a$ and the effective range $r_s$ of the
LM2M2 potential, widely used in the literature~\cite{aziz:1991_j.chem.phys.}. 
The variation of the potential strength produces different values of
the scattering length $a$ allowing a comparison of the recombination
rate given by Eq.(\ref{eq:recomb}). The model includes a three-body interaction
necessary to tune the trimer energy of the LM2M2 potential. This
procedure is equivalent, up to range corrections, to the implementation
of effective field theory (EFT) at leading order (LO).
This strategy has been used before in the
study of the atom-dimer effective range function
$ka\cot\delta$~\cite{kievsky:2013_phys.rev.a}. It has
been found that the zero-range formula
of Eq.(\ref{eq:cotdelta}) has to be modified in order to describe
results obtained through the solution of the Schr\"odinger equation using 
finite-range potentials. The modification consists in a shift on the variable 
$\kappa_*a$, and in the replacement of $ka\cot\delta$ by $ka_B\cot\delta$, where
$a_B$ is related to the two-body binding energy through $E_2=\hbar^2/ma_B^2$. 
In the present analysis we find that the same type of modification has to
be made in order to describe the numerical results.

The universal character of Eq.(\ref{eq:recomb}) allows its application to
very different systems. Here we extend the study on atomic systems
to describe a nuclear system: the neutron-neutron-proton recombination rate
close to threshold. This study is twofold; from one side, we would like to
confirm that systems whose energies and sizes differ by several order of
magnitude are still described by the same universal equation. On the other hand, 
very low energy recombination in nuclear systems can be achieved in stars, and the 
present study can therefore be applied to those systems in which the three-body 
structure is dominant, e.g., neutron-neutron-core systems. As an example we can mention the
recently performed studies in recombination of
$^3$H~\cite{deltuva:2013_arxiv:1301.1905[nucl-th]} and low energy $n-^{19}$C
collisions~\cite{yamashita:2008_physicslettersb}.

Finally, we remark that in cold-atom physics the search for universal behavior
is a very active sector of research.
At present, there is an intense experimental activity to study Efimov
physics in trapped ultracold gases. In these systems the recombination rate has a major 
importance, being the main loss mechanism. Three different values of the two-body
scattering length have relevance: $a_-$, which characterizes the threshold
where the trimer disappears into the three-atom continuum,
$a_*$, which characterizes the threshold where the trimer disappears into 
the atom-dimer continuum, and $a_+$, which characterizes a minimum in the 
recombination rate. Discrepancies are found between the zero-range
theoretical predictions and experimental
determinations of these three
quantities~\cite{zaccanti:2009_natphys,*roy:2013_phys.rev.lett.,ferlaino:2011_few-bodysyst.,dyke:2013_arxiv:1302.0281[cond-mat.quant-gas],machtey:2012_phys.rev.lett.}.
To this respect, we would like to see if the
quantities obtained from the calculations using finite range potentials
are in better agreement with the experimental predictions. 

The paper is organized as follows. In the next section the two-body
and three-body potential models are introduced and the results for
the first two energy levels are discussed. The main results of
this work are given in Section III, which is divided into four subsections
dedicated to analyze the elastic and breakup cross sections, the
recombination and dissociation rates, comparison of the results to
experimental data and nucleon-deuteron scattering
above threshold. The conclusions are given in the last section.

\section{The three-boson system} 

In this section we study universal aspects of a three-boson system by taking
the three-helium system as reference. At the two-body level, we consider
one of the most commonly used He-He potentials, i.e., the LM2M2 
interaction~\cite{aziz:1991_j.chem.phys.}, which is taken as the reference 
interaction. In particular, in order to explore the $(a^{-1},\kappa)$ plane
($\kappa={\rm sign}(E)[|E|/(\hbar^2/m)]^{1/2}]$ and $E$ the energy level), 
we modify this potential as: 
\begin{equation}
  V_\lambda(r)=\lambda V_{\text{LM2M2}}(r)\,\, .
\label{eq:mtbp}
\end{equation}

Examples of this strategy exist in the
literature~\cite{esry:1996_phys.rev.a,barletta:2001_phys.rev.a,gattobigio:2012_phys.rev.a}.
For $\lambda\approx 0.9743$ the interaction is close to the unitary limit
($a\rightarrow\infty$). For $\lambda=1$ 
the values predicted by the LM2M2 are recovered: the scattering length $a=189.41$ $a_0$, 
the two-body energy $E_2$=-1.303 mK, and the effective range $r_e=13.845$ $a_0$,
with the mass parameter $\hbar^2/m=43.281307~\text{($a_0$)}^2\,\text{K}$. 

Following
Ref.~\cite{kievsky:2011_few-bodysyst.,gattobigio:2012_phys.rev.a,kievsky:2013_phys.rev.a}
we define an attractive two-body gaussian (TBG) potential
\begin{equation}
V(r)=V_0 \,\, {\rm e}^{-r^2/r_0^2}\,,
\label{eq:twobp}
\end{equation}
with range $r_0=10$~$a_0$ and strength $V_0$ fixed to reproduce the values of $a$ given
by $V_\lambda(r)$. 
For example with the strength $V_0=-1.2343566$~K, corresponding to $\lambda=1$,
the LM2M2 low-energy data are closely reproduced, 
$E_{2}=-1.303$ mK, $a=189.42$ $a_0$, and $r_e=13.80$ $a_0$

The use of the TBG potential in the three-atom system produces a ground state binding
energy appreciable deeper than the one calculated with $V_\lambda(r)$.
For example, at $\lambda=1$ the LM2M2 helium-trimer 
ground-state-binding energy is $E_3^0=126.4$ mK, whereas the one obtained using
the two-body-soft-core potential in Eq.~(\ref{eq:twobp})
is $151.32$ mK. To solve this discrepancy we introduce a repulsive
hypercentral-three-body (H3B) interaction
\begin{equation}
W(\rho_{123})=W_0 \,\, {\rm e}^{-\rho^2_{123}/\rho^2_0}\,,
  \label{eq:hyptbf}
\end{equation}
with the strength $W_0$ tuned to reproduce the trimer energy $E_3^0$ obtained with
$V_\lambda(r)$ for all the explored values of $\lambda$.
Here $\rho^2_{123}=\frac{2}{3}(r^2_{12}+r^2_{23}+r^2_{31})$ is the hyperradius
of three identical particles and $\rho_0$ gives the range of the three-body force. 
Following Ref~\cite{gattobigio:2012_phys.rev.a} we use $\rho_0=r_0$.
It should be noticed that the description of the three-boson systems using
a two- plus three-body interaction constructed to reproduce the low energy data
is equivalent, up to finite range corrections, to a description based
on  EFT at  LO
(see Ref.~\cite{bedaque:1999_nucl.phys.a} and references therein).

Varying $\lambda$ from the unitary limit to $\lambda=1.1$ we obtain a
set of values for the ground state binding energy $E_3^0$ and first excited 
state $E_3^1$ using the TBG and TBG+H3B potentials in a broad range of $a$.
The results can be compared to the predictions given by the
Efimov's binding energy equations 
\begin{equation}
 E_3^n+\frac{\hbar^2}{ma^2}= \text{e}^{-2(n-n^*)\pi/s_0}
\exp{[\Delta(\xi)/s_0]}\frac{\hbar^2\kappa_*^2}{m} \,,
\label{eq:uni3}
\end{equation}
where $\tan\xi=-(mE_3^n/\hbar^2)^{1/2}a$ and the function $\Delta(\xi)$ can be
found in \cite{braaten:2006_physicsreports}.
Fixing $n^*=1$, the three-body parameters $\kappa_*$ is determined by
calculating $E_3^1$ at
the unitary limit; we obtain $\kappa_*=2.119\times 10^{-3}$~$a_0^{-1}$ and
$\kappa_*=1.899\times 10^{-3}~a_0^{-1}$ for the TBG and TBG+H3B, respectively. 

\begin{figure}
  \includegraphics[width=\linewidth]{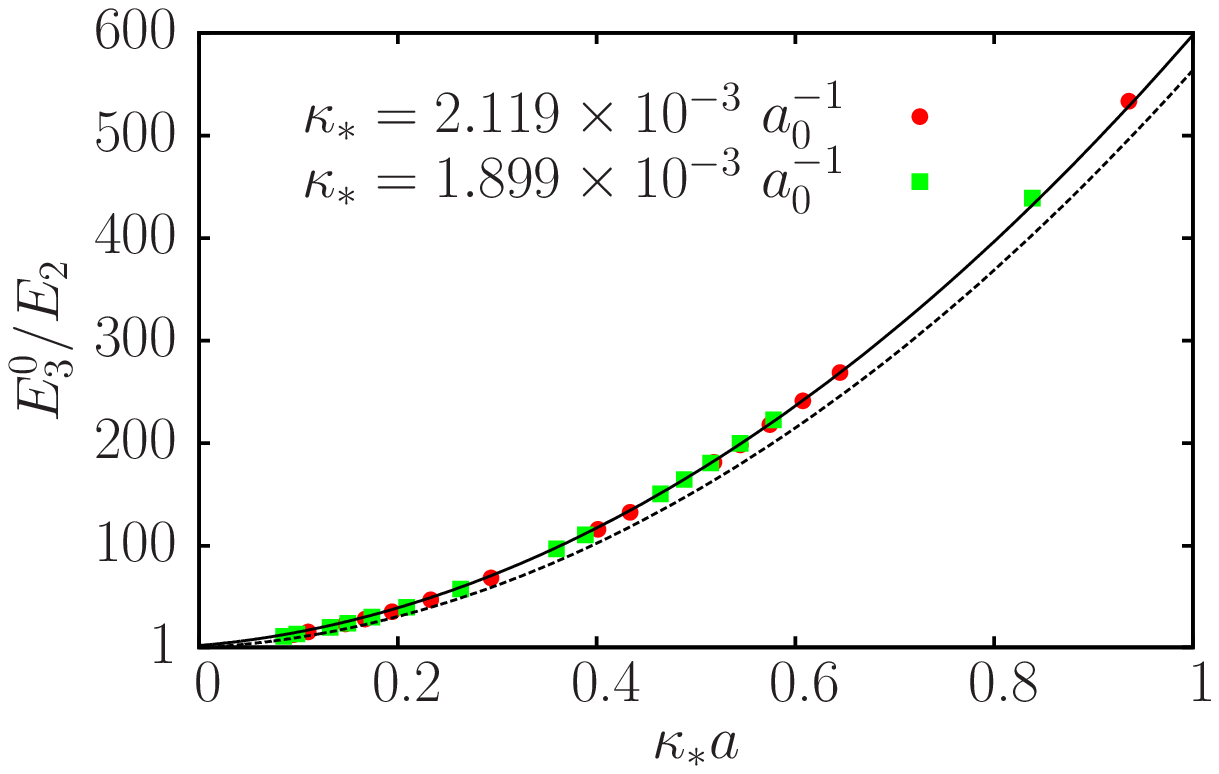}
  \includegraphics[width=\linewidth]{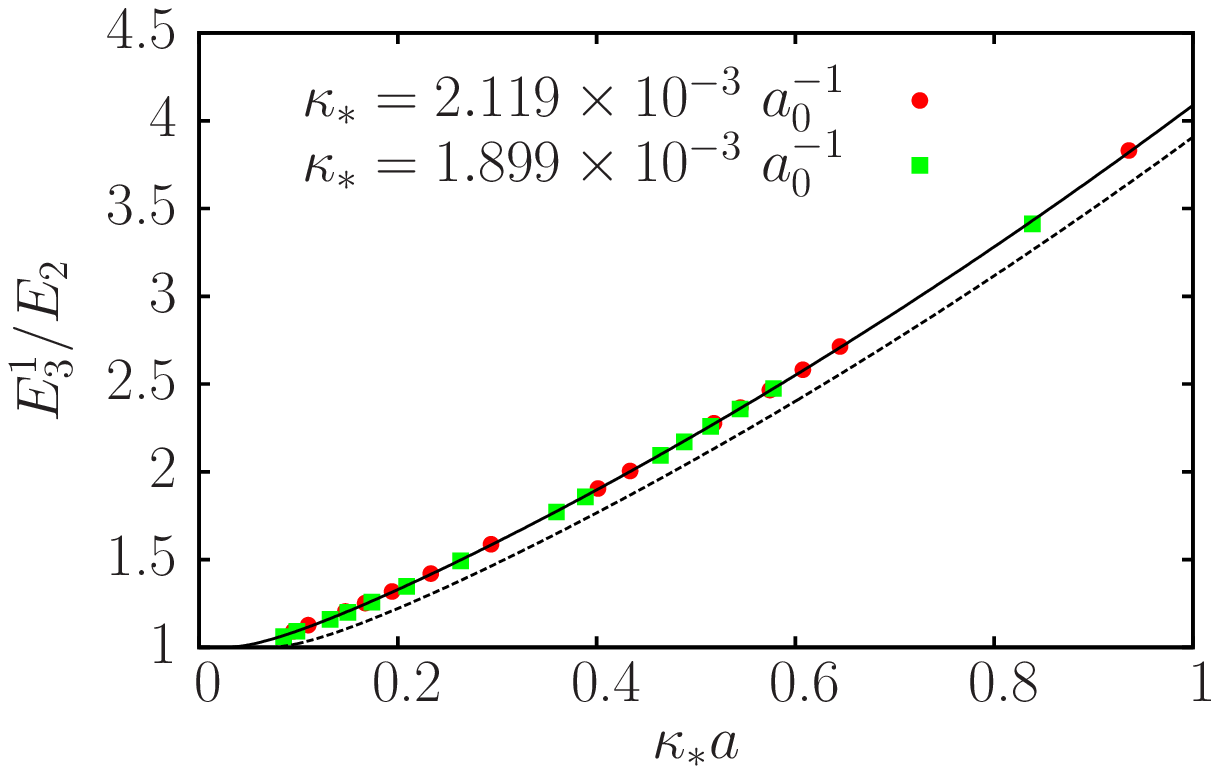}
  \caption{(Color online) Ratio between the energy of the ground (upper panel)
    and first excited (lower panel) state of the trimer and the dimer binding
    energy as a function of $\kappa_*a$.  The dashed line is the universal
    prediction of the Efimov law  Eq.(\ref{eq:uni31}) without shift
  ($\Gamma_n=0$), while the solid line is the translated universal curve.  The
  circles and squares are the calculations using the TBG and TBG+H3B potentials
  respectively.}
\label{fig:plotnew}
\end{figure}

It has been shown in
Ref.~\cite{kievsky:2013_phys.rev.a} that in order to be
in accord with the numerical results obtained solving a finite-range potential,
the universal relation Eq.~(\ref{eq:uni3}) must be modified in the following way
\begin{equation}
   E_3^n/E_2=\tan^2\xi\,,\quad
   \kappa_*a= \text{e}^{(n-n^*)\pi/s_0} \exp{[-\Delta(\xi)/2s_0]}/\cos\xi  -
   \Gamma_n\,,
  \label{eq:uni31}
\end{equation}
where the finite-range nature of the interaction has been taken into account by the
substitution $\hbar^2/ma^2\rightarrow E_2$, and by the shift $\Gamma_n$.  In
Fig.~\ref{fig:plotnew} we collect our numerical results for the ratios
$E_3^0/E_2$ and $E_3^1/E_2$ as a function of $\kappa_*a$ for the TBG potential
(circles) and of the TBG+H3B potential (squares).  In the upper panel of
Fig.\ref{fig:plotnew} we report the calculations for the ground state. The
dashed line corresponds to Eq.~(\ref{eq:uni31}) without shift, while the solid
line, which fits our numerical results, has
$\Gamma_0 \simeq 3\times 10^{-2}$. In the lower panel of
Fig.~\ref{fig:plotnew} we report the calculations for the excited state. 
As above, the dashed line corresponds to Eq.~(\ref{eq:uni31}) without shift,
while the solid has a finite shift, $\Gamma_1 \simeq 4\times 10^{-2}$.

A more accurate analysis of the numerical results reveals that the shift depends
on  $\kappa_*$~\cite{kievsky:2013_phys.rev.a}; 
in first approximation, for the excited state, 
we can write $\Gamma_1 \simeq \kappa_*r_*$ with $r_*\approx 21~a_0$.
From Eq.(\ref{eq:uni31}) we can extract $a_*^1$, the scattering length at
which the excited state $E_3^1$ disappears into the atom-dimer continuum.
This happens when $E_3^1/E_2=1$ or the angle $\xi=-\pi/4$. Using the very accurate
result, $\Delta(-\pi/4)=6.0273$, given in Ref.~\cite{braaten:2006_physicsreports}, 
we obtain the relation
\begin{equation}
   \kappa_*a^1_*= 0.07076 -\Gamma_1
  \label{eq:astar}
\end{equation}
which in our case gives $a_*^1\approx 14.5~a_0$ and $a_*^1\approx 16.2~a_0$
for the TBG and TBG+H3B potentials respectively. Moreover, dividing the above 
relation by $\kappa_*$, we can write $r_*=a^1_{*,zr}-a^1_*$, where we have 
introduced the universal zero-range scattering length $a^1_{*,zr}=0.07076/\kappa_*$.
From this relation we can interpret the shift, in units of
$\kappa_*^{-1}$, as the difference between the zero-range and the finite-range
predictions for $a_*$. In addition, the relation applied to the $n=1$ branch can
be extended to the different $n$-branches. 

\section{atom-dimer scattering above the breakup threshold.} 

In Ref.~\cite{kievsky:2013_phys.rev.a} the universal character of
the atom-dimer scattering below the breakup threshold has been discussed
in terms of the potential model introduced in the previous section.
It has been shown that when finite-range interactions are used, the universal formula of 
the atom-dimer scattering length given in Eq.(\ref{eq:a_AD}) can be modified as
\begin{equation}
  a_{AD}/a_B=d_1 +d_2\tan[s_0\ln(\kappa_*a + \Gamma_*)+d_3]\,,
\label{eq:ab_AD}
\end{equation}
where $a_B$ is defined from the relation $E_2=\hbar^2/ma_B^2$, 
and $\Gamma_*\simeq\Gamma_1$ is the shift for the atom-dimer scattering

Moreover, the effective range function of Eq.(\ref{eq:cotdelta})
can be adapted to describe finite-range interactions too. In particular, 
it is given by
\begin{equation}
  ka_B\cot\delta=c_1(ka) +c_2(ka)\cot[s_0\ln(\kappa_*a + \Gamma_{\text{e}})+\phi(ka)]\,,
  \label{eq:abcotdelta}
\end{equation}
with the effective-range shift
$\Gamma_{\text{e}}\simeq 4\times 10^{-2}$~\cite{kievsky:2013_phys.rev.a}.
This modified equation agrees with the numerical calculations 
of Ref.~\cite{kievsky:2013_phys.rev.a}, which have been
performed by using the potential models of the previous section
and the hyperspherical harmonic (HH) method in conjunction 
with the Kohn variational principle~\cite{kievsky:1997_nuclearphysicsa}, for a
wide range of values of $\kappa_*a$ varying from $0.26$
to $0.94$. In this range the effective range function presents different
patterns as a function of the energy. For the lowest values of $\kappa_*a$
it was almost linear whereas when increasing $\kappa_*a$ a pole structure appeared
(see Fig.4 of Ref.~\cite{kievsky:2013_phys.rev.a}).
In addition, it was shown that around the value $\kappa_*a\approx 0.54$
the structure of the effective range function coincides with the one describing
neutron-deuteron scattering. In this way, a confirmation of the
universal character of Eqs.(\ref{eq:ab_AD}) and (\ref{eq:abcotdelta}) has been done
for systems with very different typical lengths.

The fact that $\Gamma_e\approx\Gamma_*\approx\Gamma_1$ can be understood
noticing that, for the values of $a$ considered, $E_3^1$ is the only excited
state of the three-boson system. Probably, for larger values of $a$, when
a second excited state appears, a different shift has to be considered.
The analysis of atom-dimer scattering for larger values of $a$ is a numerical
difficult task and remains outside the scope of the present work.  
From Eq.(\ref{eq:ab_AD}) it is possible to extract the value of $a^1_*$ by
equating the argument of the tangent to $-\pi/2$. Using the value of
$d_3=1.100$ given in Ref.~\cite{kievsky:2013_phys.rev.a},
the results are $a_*^1\approx 14.3~a_0$ and $a_*^1\approx 16.0~a_0$
for the TBG and TBG+H3B potentials respectively, in complete agreement with
those obtained using Eq.(\ref{eq:uni31}). These values can be used to evaluate
the ratio $a_*^1/a_-^0$ that appears frequently in the literature from measurements
on trapped ultracold atoms~\cite{zaccanti:2009_natphys,ferlaino:2011_few-bodysyst.,
machtey:2012_phys.rev.lett.}. The universal zero-range theory predicts
$a_*^1/a_-^0\approx -1.06$ whereas using the values $a^0_-\approx -43.3~a_0$ and
$a^0_-\approx -48.1~a_0$ for the TBG and TBG+H3B potentials respectively,
given in Ref.~\cite{gattobigio:2012_phys.rev.a}, 
we obtain $a_*^1/a_-^0\approx -0.32$ in both cases. An analysis of the present
result in comparison to those given by different experimental groups is 
given in the Section III C.

\subsection{Elastic and breakup cross sections}
The universal character of Eqs.(\ref{eq:ab_AD}) and (\ref{eq:abcotdelta})
has been deeply studied in Ref.~\cite{kievsky:2013_phys.rev.a}
in a large range of $a$ values. Here we extend the analysis of atom-dimer
scattering to energies above
the breakup threshold. In particular, we study the universal
form predicted by Petrov~\cite{petrov:2005_} for the recombination rate at threshold.
To this aim we make use of the adiabatic expansion method as discussed
in Refs.~\cite{kievsky:2011_few-bodysyst,romero-redondo:2011_phys.rev.a} for
energies below the dimer breakup threshold and recently extended to
energies above that threshold~\cite{garrido:2012_phys.rev.a}. As it is well known, the
adiabatic expansion method is a very powerful method used to describe bound
states~\cite{nielsen:2001_phys.rep.}. However, the extension to describe scattering
states encountered some difficulties, in particular in the case of atom-dimer
elastic scattering. The problem arose from the difference between the set
of coordinates in which the process has a natural asymptotic description (the
usual Jacobi coordinates) and the expansion in terms of hyperradial functions which produced
a low rate of convergence. In order to circumvent the problem, two
integral relations have been derived in Ref.~\cite{barletta:2009_phys.rev.lett.} and already 
applied several times in the
literature~\cite{kievsky:2011_few-bodysyst,kievsky:2011_few-bodysyst.,nollett:2012_phys.rev.c}.
Essentially the method establishes that the scattering matrix is obtained from
the following two matrices
\begin{eqnarray}
        B_{ij} & = &  \frac{2m}{\hbar^2} 
   \langle\Psi_i^t | \hat{\cal H}-E |F_j \rangle \\
        A_{ij} & = & - \frac{2m}{\hbar^2}  
  \langle \Psi_i^t|\hat{\cal H}-E |G_j \rangle 
 \label{eq:intrel},  
\end{eqnarray}
where the indexes $i,j$ label the ingoing and outgoing channels (either
elastic or inelastic), $\Psi_i^t$ is the solution of the adiabatic equations
at a given energy $E$, and $F_j,G_j$ are the ingoing and outgoing 
solutions of the free Schr\"odinger equation $(T-E)F,G=0$  
(see Ref.~\cite{garrido:2012_phys.rev.a} for details). The ${\cal S}$-matrix
is then given by the product $A^{-1}B$. The integral relations have a short-range 
character, and the problem mentioned above of the mismatch of the coordinates 
has not consequences once the internal part of the scattering wave function is
properly described. 

For energies above the dimer breakup threshold,
the unitarity of the ${\cal S}$-matrix implies that given an incoming channel,
for instance channel 1 (1+2 channel), we have that
\begin{equation}
\sum_{n=2}^\infty |{\cal S}_{1n}|^2=1-|{\cal S}_{11}|^2,
\end{equation}
which means that computation of the elastic term ${\cal S}_{11}$
amounts to computation of the infinite summation of the $|{\cal S}_{1n}|^2$ 
terms ($n>1$) corresponding to the breakup channels.
The complex value of ${\cal S}_{11}$ can be written in terms of a complex
phase-shift $\delta$ as:
\begin{equation}
{\cal S}_{11}=e^{2i\delta}=e^{-2 \mbox{\scriptsize Im}(\delta)} e^{2i
\mbox{\scriptsize Re}(\delta)}
=|{\cal S}_{11}|e^{2i \mbox{\scriptsize Re}(\delta)}.
\label{eq22}
\end{equation}
The value of $|{\cal S}_{11}|^2$ gives the probability of elastic
atom-dimer scattering, and
$|{\cal S}_{11}|$ is usually referred to as the inelasticity parameter
(for example denoted by $\eta$ in
\cite{friar:1990_phys.rev.c,friar:1995_phys.rev.c,kievsky:2001_phys.rev.c,kievsky:1999_phys.rev.lett.}). 
Obviously, the closer the inelasticity to 1 the more
elastic the reaction.
In fact, for energies below the breakup threshold the phase-shift is real and
$|{\cal S}_{11}|=1$.

\begin{figure}
  \includegraphics[width=\linewidth]{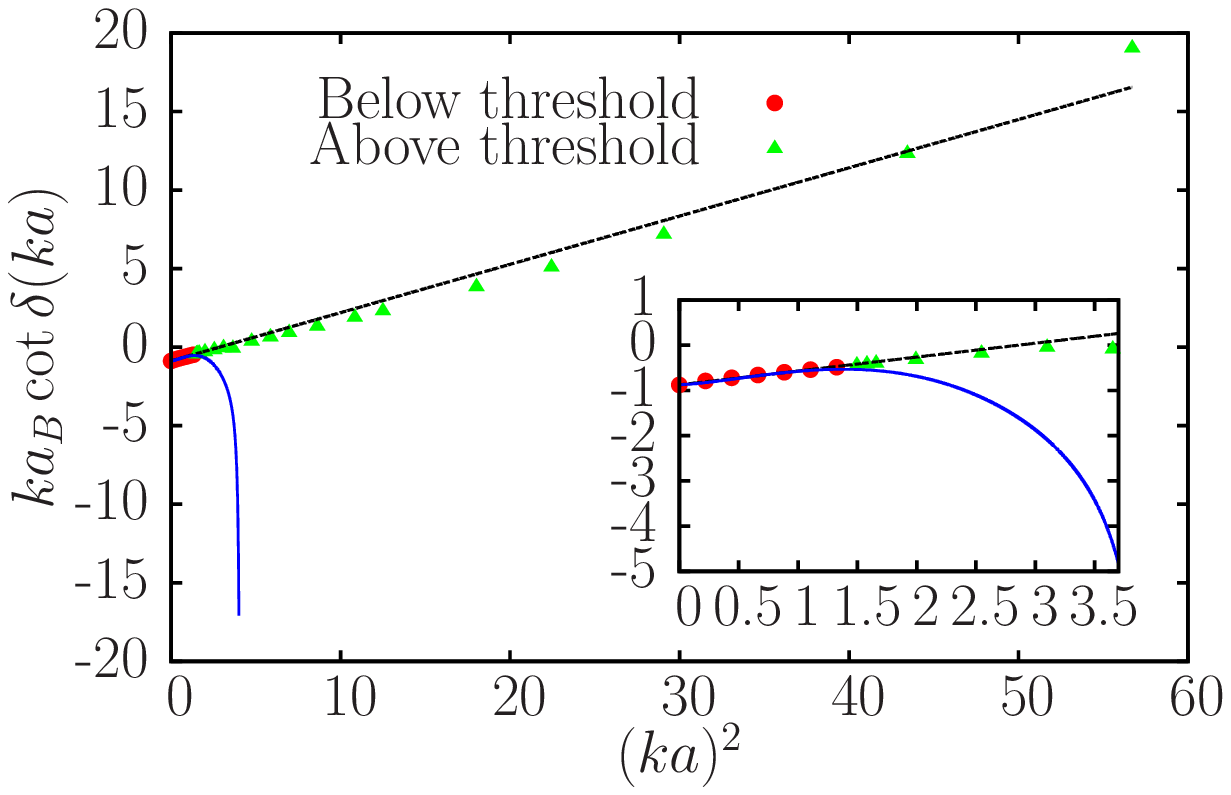}
  \includegraphics[width=\linewidth]{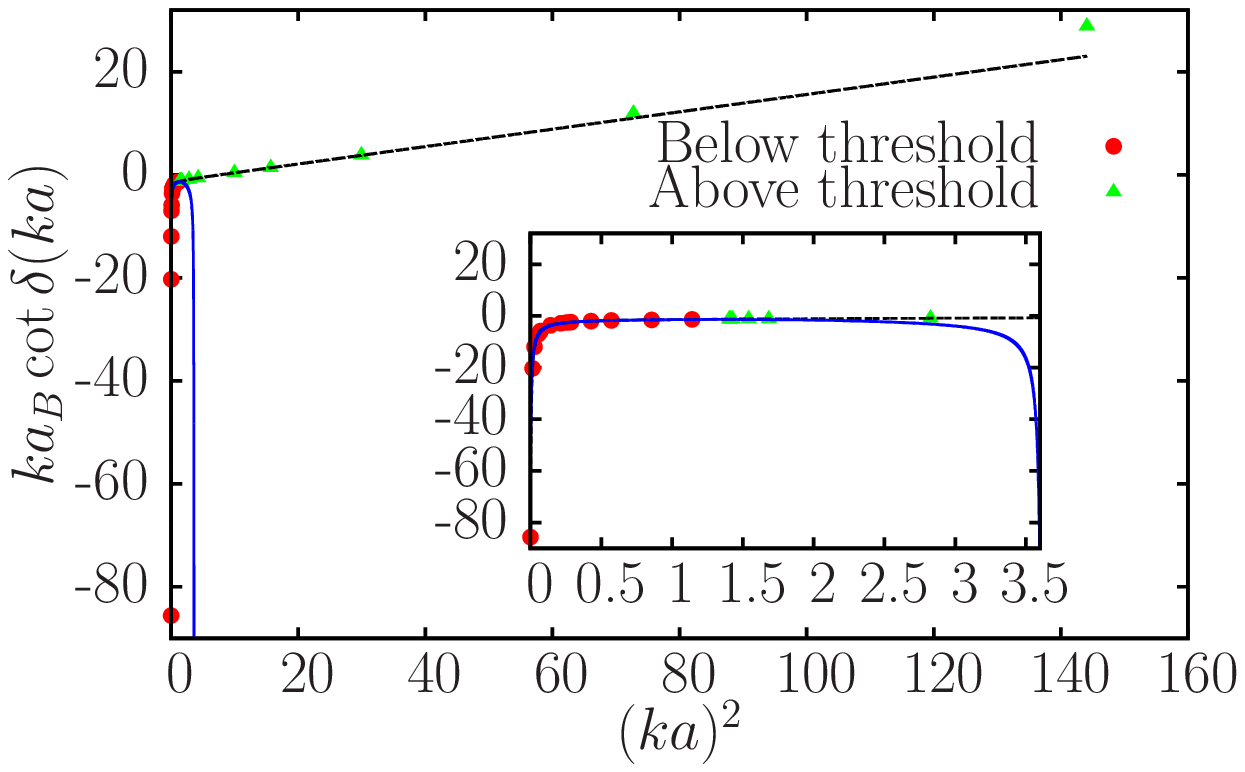}
  \caption{(Color online) The effective range function as a function
of $(ka)^2$ for two different values of $\kappa_*a$; in the upper panel
$\kappa_*a = 0.36$, and in the lower panel $\kappa_*a = 0.56$. 
The dimer threshold corresponds to $(ka)^2\approx 4/3$. 
The (red) circles are the calculations below the dimer threshold of 
Ref.~\cite{kievsky:2013_phys.rev.a}. The (green)
triangles are the present calculations. The solid line is
Eq.(\ref{eq:abcotdelta}) whereas the dashed line
is an effective range parametrization (see text).}  
\label{fig:redelta}
\end{figure}

The convergence pattern of the method has been studied in 
Ref.~\cite{garrido:2012_phys.rev.a}, where the inelasticity 
($|{\cal S}_{11}|$) and the real part of the phase-shift 
(Re($\delta$)) have been calculated at selected energies
for increasing values of $K_{\mbox{\scriptsize max}}$ (grand-angular quantum number 
associated to the last adiabatic term included in
the expansion of the scattering wave function). The conclusion was that
the use of the integral relations produces a pattern of convergence similar
to a bound-state calculation. For example, a $K_{\mbox{\scriptsize max}}$ 
value of around 12 is enough to get a rather well converged inelasticity, 
while Re($\delta$) requires a few more adiabatic terms in order to reach 
convergence. 

To make contact with the results in
Ref.~\cite{kievsky:2013_phys.rev.a}, in
Fig.~\ref{fig:redelta} we show the effective range function of the TBG potential
corresponding to $\kappa_*a=0.36$ (upper panel), and to $\kappa_*a=0.56$
(lower panel) for a wide range of energies.
For
energies below the dimer threshold the circles (red online) are the results of
Ref.~\cite{kievsky:2013_phys.rev.a}, whereas for
energies above the dimer threshold, our results obtained using Re($\delta$) in
the definition of the effective range function are given by the triangles (green
online). 
The solid line has been obtained applying Eq.(\ref{eq:abcotdelta}), and
the dashed line is the effective range parametrization of the low energy points;
in the upper panel, the effective-range is 
$k\cot\delta\approx -1/a_{AD}+1/2 \,r_{\text{eff}} \,k^2$, with
$r_{\text{eff}}$ used as a parameter. In the lower panel, a pole-structure 
dominates the low-energy behaviour of the effective-range function, which is 
better parametrized by $k\cot\delta\approx (-1/a_{AD}+1/2\,r_{\text{eff}} \,k^2 -
P\,r_{\text{eff}}^3 \,k^4)/(1+k^2/k_0^2)$; in addition to the effective
range we must introduce the shape parameter $P$ and the momentum of the pole
$k_0$.
From the figure we can observe that the
universal form of Eq.(\ref{eq:abcotdelta}) does not describe the points above
the dimer threshold. On the other hand, the effective range parametrization
remains close to the computed values in a larger range. 

\begin{figure}
\begin{center}
  \includegraphics[width=\linewidth]{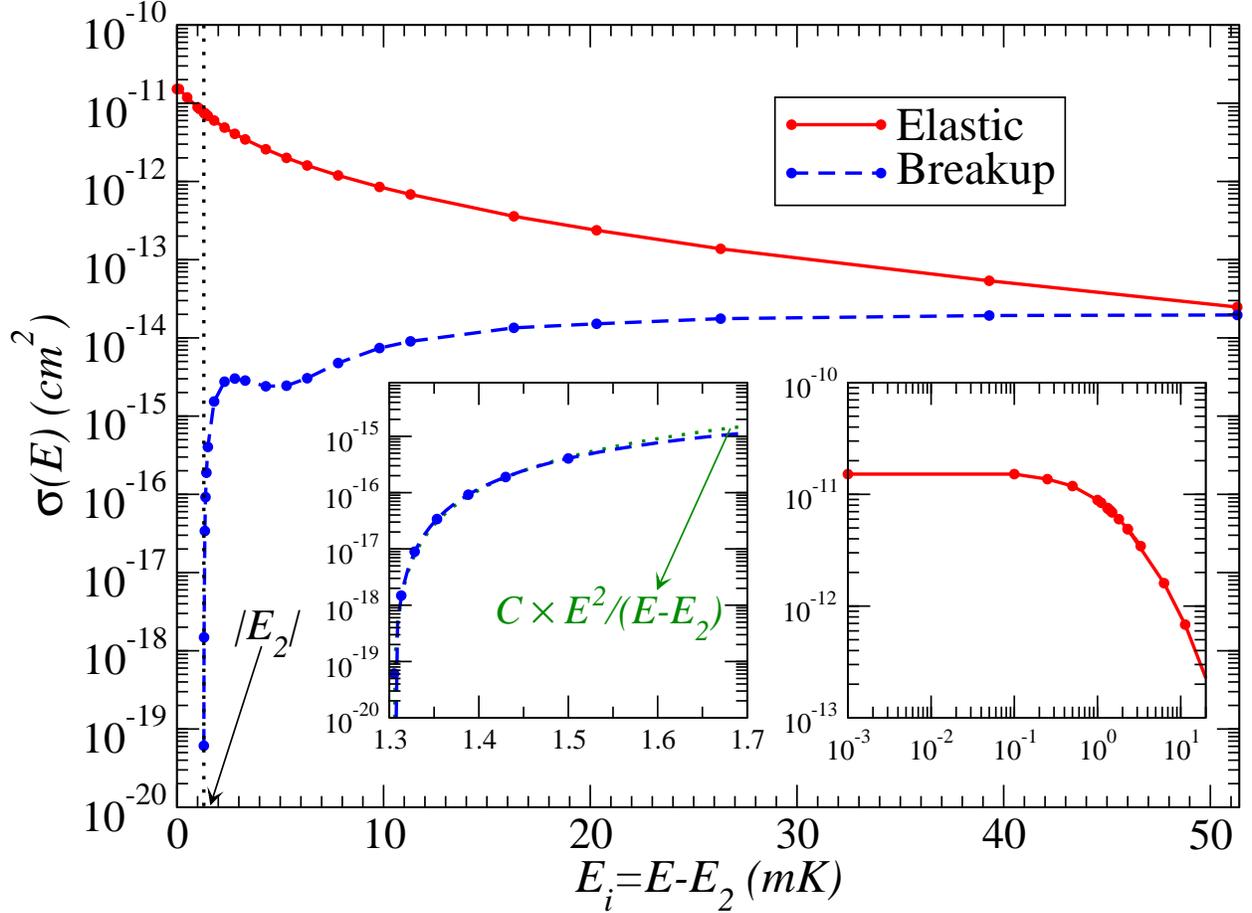}
  \caption{(Color online) Atom-dimer elastic (solid) and breakup (dashed) cross sections 
as a function of the atom incident energy for a three-helium atom system and the TBG+H3B
potential. The breakup channel is open at $E_i=E_2=1.303$ mK. The
zoom shows the behavior close to both thresholds, i.e., $E_i=0$ for the
elastic case  and $E_i=E_2$ for the breakup case. The circles correspond to
actual calculations, while the curves are interpolations.}
\label{fig:sigmas}
\end{center}
\end{figure}

Limiting the discussion to the $L=0$ channel, the determination of the elastic
matrix element ${\cal S}_{11}$ permits the computation of the elastic
($\sigma_e$) and breakup ($\sigma_b$) cross sections, which are given by the well-known expressions:
\begin{eqnarray}
&\displaystyle\sigma_e =\frac{\pi}{k^2}|1- {\cal S}_{11}|^2 \\
&\displaystyle\sigma_b =\frac{\pi}{k^2}(1- |{\cal S}_{11}|^2)
\label{eq:sigma}
\end{eqnarray}
with $k^2=(4/3)E_i/(\hbar^2/m)$ and $E=E_i+E_2$, the total energy of the process.

Following the method described in \cite{garrido:2012_phys.rev.a}, we have
computed $\sigma_e$ and $\sigma_b$ for the TBG+H3B potential ($\lambda=1$ case).
The results are given by the solid and dashed curves in Fig.~\ref{fig:sigmas},
respectively, where the cross sections are shown for incident energies 
$E_i$ up to $50$ mK. At this energy 
both cross sections have a similar size. In the figure the corresponding behavior at threshold are zoomed.
In the case of the elastic scattering, when $E_i$ approaches 0 the cross 
section reaches the constant value of $\sigma_e\rightarrow 4 \pi a^2_{AD}$.
In the case of the breakup cross section the threshold corresponds
to $E=0$ (or $E_i=E_2$), and we have that for $E\rightarrow 0$ the breakup cross
section $\sigma_b$ behaves as $\sigma_b\propto E^2/(E-E_2)$ 
(or, in other words, $1-|{\cal S}_{11}|^2 \propto E^2$).

\subsection{Recombination and dissociation rates}

The breakup cross section described in the previous subsection is directly
related to the dissociation rate, $D_3$, for the $\mbox{$^4$He}_2+\mbox{$^4$He}
\rightarrow \mbox{$^4$He}+\mbox{$^4$He}+\mbox{$^4$He}$ process.  In particular,
for three identical bosons with angular momentum and parity $0^+$ it takes the 
form \cite{suno:2008_phys.rev.a}:
\begin{equation}
D_3=\hbar \frac{k}{\mu_{A,d}}\sigma_b
=\frac{\hbar \pi}{\mu_{A,d} k} \left( 1 - |{\cal S}_{11}|^2\right),
\end{equation}
where $\mu_{A,d}=(2/3)m$ is the atom-dimer reduced mass and $m$ is the mass of
the atom.

Also, making use of the detailed balance principle, it is possible to relate
$\sigma_b$ to the cross section corresponding to the inverse process
$\mbox{$^4$He}+\mbox{$^4$He}+\mbox{$^4$He}\rightarrow
\mbox{$^4$He}_2+\mbox{$^4$He}$. This permits to obtain the recombination 
rate, $K_3$, for such process in terms of the inelasticity parameter. 
In particular, again for three identical bosons with angular momentum 
and parity $0^+$, $K_3$ takes the form \cite{suno:2008_phys.rev.a}:
\begin{equation}
 K_3=3! \frac{32\hbar \pi^2}{\mu k_3^4} \left( 1 - |{\cal S}_{11}|^2\right) 
\label{recom}
\end{equation}
where $\mu^2=m^2/3$ and $k_3^2=2E/(\hbar^2/\mu)$.

\begin{figure}
\begin{center}
  \includegraphics[width=\linewidth]{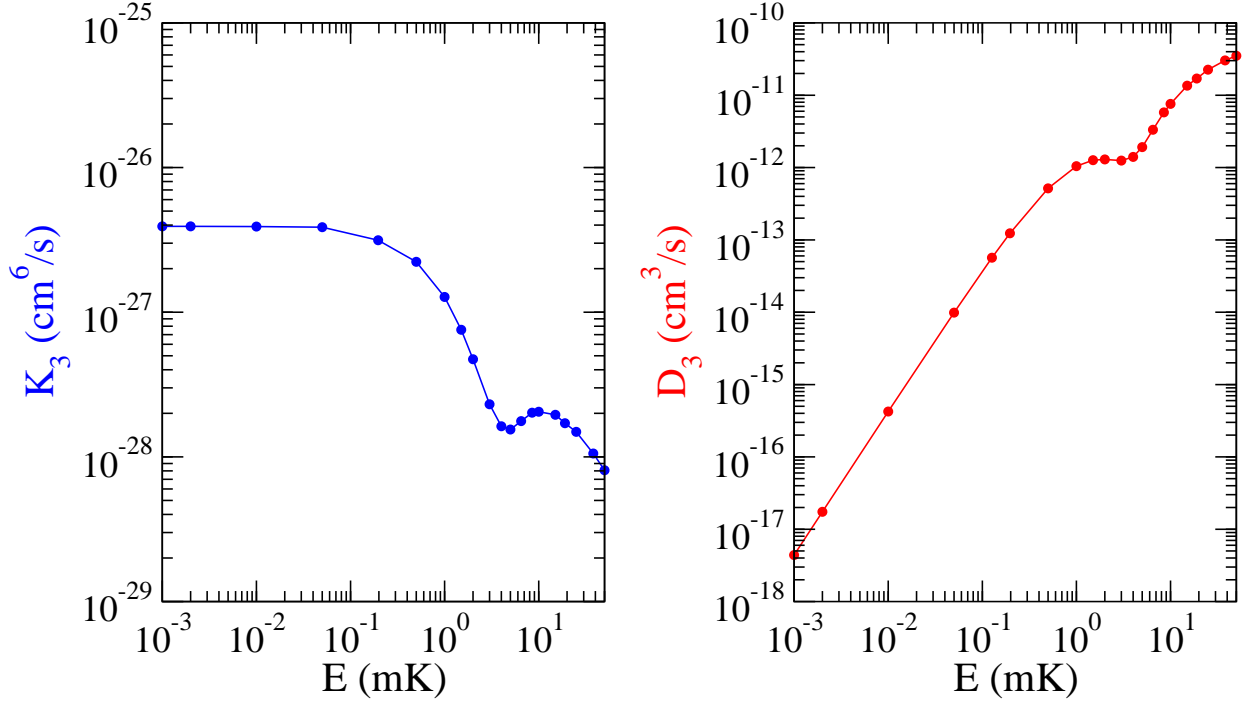}
  \caption{(Color online) The recombination rate $K_3$ and the dissociation rate
  $D_3$ for three helium atoms with the TBG+H3B potential ($\lambda=1$) as a
  function of the three-body energy $E$.}
\label{fig:recom}
\end{center}
\end{figure}

The $K_3$ behavior at low
energies results proportional to $E^{K_m}$ as demonstrated in
Ref.~\cite{esry:2001_phys.rev.a} ($K_m$ is the smallest grand-angular quantum number
associated with the continuum adiabatic channels). In the present case
$K_m=0$ and $K_3$ is almost constant as $E \rightarrow 0$. In the case
of $D_3$, its low energy behavior follows the $E^{K_m+2}$ rule derived in
Ref.~\cite{esry:2001_phys.rev.a}. These behaviors can be seen in Fig.\ref{fig:recom}
in which $K_3$ and $D_3$ are displayed as a function of the three-body energy $E$
for the TBG+H3B calculation ($\lambda=1$ case). This low-energy behavior can
also be deduced from the behavior of $\sigma_b$ close to threshold 
shown in the zoom of Fig.~\ref{fig:sigmas}, and in particular from the
fact that when the total energy goes to zero we have that $1 - |{\cal S}_{11}|^2
\propto k_3^4 \propto E^2$.

\begin{figure}
\begin{center}
  \includegraphics[width=\linewidth]{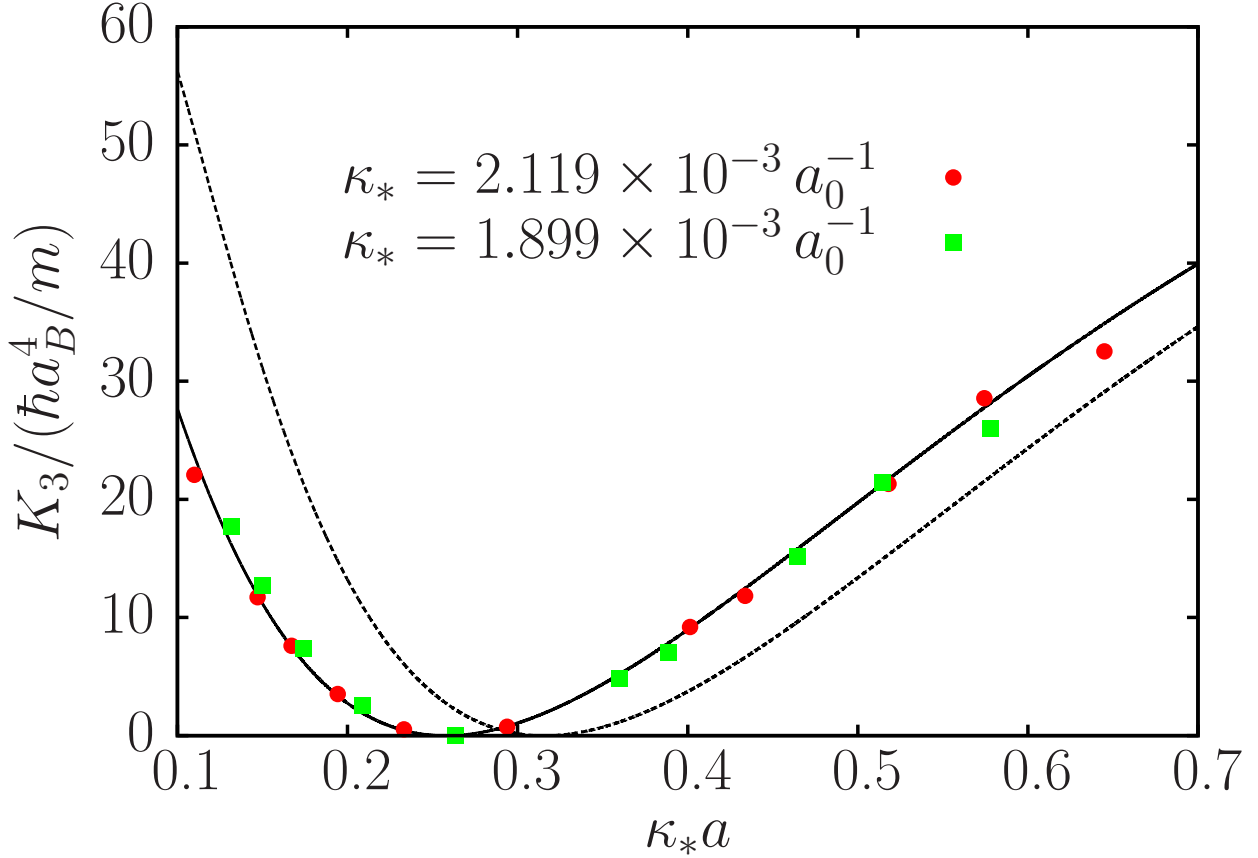}
  \caption{(Color online) The recombination rate $K_3$ at threshold
for different values of the product $\kappa_*a$. The 
circles (red online) are the results using the TBG potential whereas the
squares (green online) are the results using the TBG+H3B potential. The
points have been fitted with
Eq.(\ref{eq:abrecom}) obtaining $\Gamma_+\simeq 6\times 10^{-2}$.  }
\label{fig:recomth}
\end{center}
\end{figure}

The recombination rate at threshold can be defined as $K_3= \alpha (\hbar
a^4/m)$. In ref.~\cite{petrov:2005_} Petrov derived a log-periodic function for
$\alpha$ given in Eq.(\ref{eq:recomb}), 
whose simplified version is given by Eq.~(\ref{eq:recombSine}).
In order to use this formula to
describe our numerical results, which are obtained using finite-range
potentials, we introduce the following modification in the definition of $K_3$:
\begin{equation}
K_3=\frac{128\pi^2(4\pi-3\sqrt{3})}{\sinh^2(\pi s_0)+\cosh^2(\pi s_0)
       \cot^2[s_0\ln(\kappa_*a + \Gamma_+)+\gamma]}\frac{\hbar a_B^4}{m}
\label{eq:abrecom}
\end{equation}
with the simplified form
\begin{equation}
  K_3\approx 67.1\sin^2[s_0\ln(\kappa_*a + \Gamma_+)+\gamma] (\hbar a_B^4/m)
  \,,
  \label{eq:abrecombSine}
\end{equation}
where, as in the effective range function, we have replaced $a$ by $a_B$ and we
have introduced the shift $\Gamma_+$ in the variable $\kappa_*a$.

Our results are shown in Fig.~\ref{fig:recomth} as circles 
(TBG potential) and squares (TBG+H3B potential). The dashed line
represents the universal function of Eq.(\ref{eq:recomb})
with the value $\gamma=1.16$ from Ref.~\cite{braaten:2006_physicsreports}. It is 
interesting to see that the calculated points organize in a curve shifted 
with respect to the universal curve. The solid line represents
Eq.(\ref{eq:abrecom}), with the same value of $\gamma$ and
$\Gamma_+\simeq  6\times 10^{-2}$. By equating the argument of the sine
to zero we can extract $a^1_+$, the value of $a$ at which the recombination
rate has a minimum. We obtain the relation
\begin{equation}
   \kappa_*a^1_+=0.31575-\Gamma_+
\label{eq:apiu}
\end{equation}
which in our case results $a^1_+\approx 121~a_0$ and $a^1_+\approx 135~a_0$ for the
TBG and TBG+H3B potentials respectively. With the values $a^1_-\approx -752~a_0$ and
$a^1_-\approx -975~a_0$ for the TBG and TBG+H3B potentials respectively,
given in Ref.~\cite{gattobigio:2012_phys.rev.a}, 
we obtain $a_-^1/a_+^1\approx -6.3$ and $a_-^1/a_+^1\approx -7.2$ to be compared
to the theoretical prediction of -4.9. Moreover, defining $\Gamma_+=\kappa_*r_+$
and dividing the above relation by $\kappa_*$, we can write $r_+=a^1_{+,zr}-a^1_+$, 
where we have introduced the universal zero-range scattering length 
$a^1_{+,zr}=0.31575/\kappa_*$. In the present case we obtain $r_+\approx 29~a_0$.
As in the case of atom-dimer scattering, we can interpret the shift, in units of
$\kappa_*^{-1}$ as the difference between the zero-range and the finite-range
predictions for $a_+$. Moreover, the relation applied above to the $n=1$ branch
can be extended to the other branches.

\begin{figure}
\begin{center}
  \includegraphics[width=\linewidth]{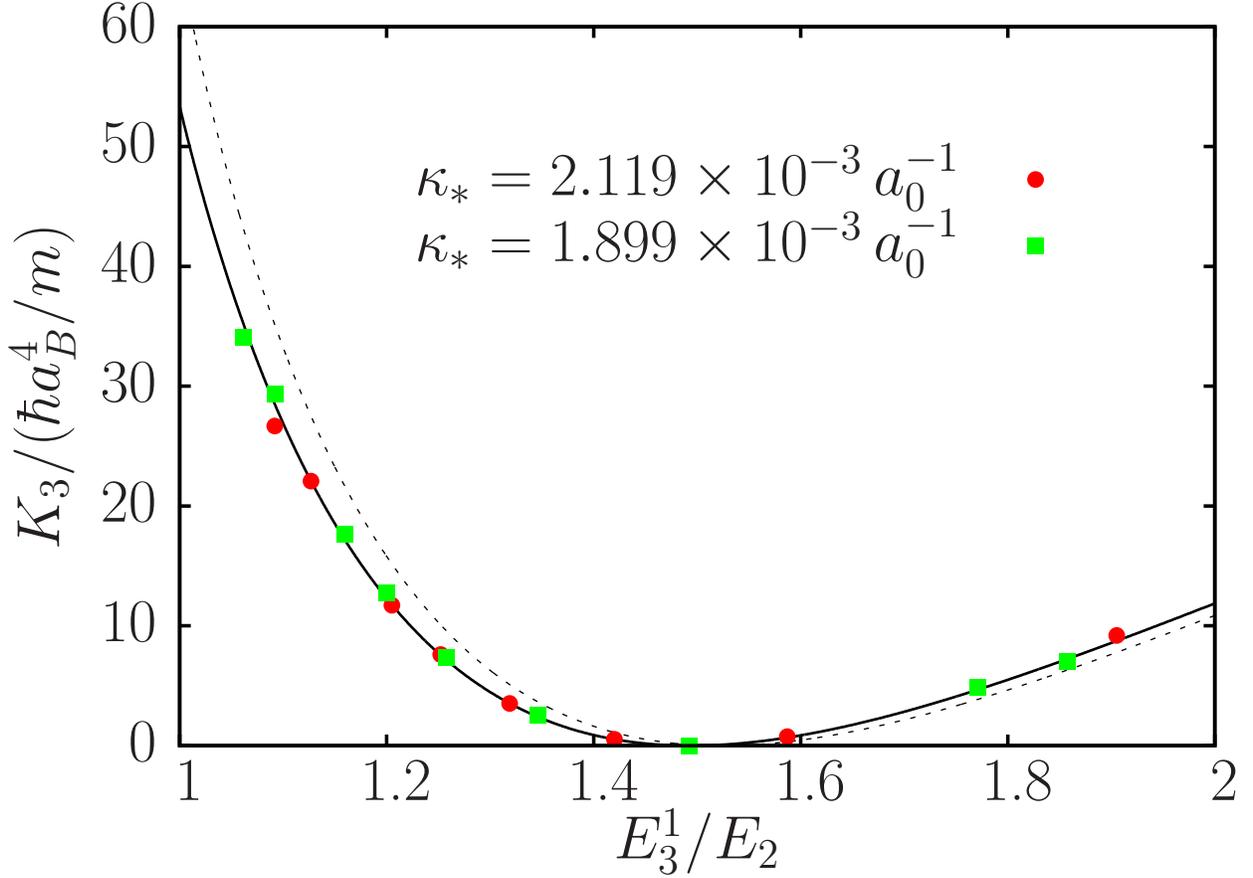}
  \caption{(Color online) The recombination rate $K_3$ at threshold
as a function of $E_3^1/E_2$. The 
circles (red online) are the results using the TBG potential whereas the
squares (green online) are the results using the TBG+H3B potential. 
The dashed line corresponds to the zero-range theory, while the solid line is
the translation of the zero-range curve.}
\label{fig:alphaEratio}
\end{center}
\end{figure}

Finally, in Fig.~\ref{fig:alphaEratio} we report our results in the form of a
Phillips plot, as has usually been done in the literature to study the
correlation between different three-body observables (see for instance
Refs.~\cite{braaten:2006_physicsreports,kievsky:2013_phys.rev.a,yamashita:2003_phys.rev.a}).
In addition to our numerical data, circles for TBG potential and squares for
TBG+H3B potential, we report the zero-range calculation without (dashed line)
and with (solid line) translation.  We observe that the translated curve fits
quite well the numerical data.  In fact, Fig.~\ref{fig:alphaEratio} can be
understood as the combination of Fig.~\ref{fig:plotnew}, lower panel, and
Fig.~\ref{fig:recomth}, where the parameter $\kappa_*a$ has been eliminated. We
have shown that the finite-range effects manifest as a translation (the shifts)
in the parameter  $\kappa_*a$, see for instance Eqs.~(\ref{eq:uni31}),
(\ref{eq:ab_AD}), (\ref{eq:abcotdelta}), (\ref{eq:abrecom}), and
(\ref{eq:abrecombSine}). Now, the smallness of the shifts justifies the fact
that the curve resulting from the elimination of the parameter $\kappa_*a$ is
itself a translated curve, at least at the first order.  If the shifts
$\Gamma_1$ and $\Gamma_+$ were  the same, the numerical calculations would had
fallen over the zero-range curve without translation. This has been for instance
demonstrated in Ref.~\cite{kievsky:2013_phys.rev.a} in the case of the
atom-dimer scattering length as a function of the excited three-body energy
(cfr. Fig.~2 of Ref.~\cite{kievsky:2013_phys.rev.a}).

\subsection{Comparison to the experimental results}

In the following we compare our numerical results using the TBG+H3B
to selected results given in the literature by different groups. 
We concentrate in the results given for $a_-$, $a_*$ and $a_+$ extracted
from resonances observed in trapped ultracold atoms. The value of $a_-$ is extracted
from a maximum in the recombination rate in the negative sector of $a$
whereas $a_*$ corresponds to a maximum in an atom-dimer resonance, and
$a_+$ to a recombination minimum in the positive sector of $a$.
In Table~\ref{tab:expres} we collect selected experimental results of these
quantities to be compared to
the results obtained using the TBG+H3B potential given in the last column.
As the numerical results, the measured ratios do not completely agree
with the zero-range universal theory. This disagreement
can be traced back to the shift introduced in the zero-range theory produced
by the finite range nature of the potential. This shift which it is not
universal produces differences between the different atomic species. However, in 
all cases the ratio $a^0_-/\ell_{vdW}\approx -9$, 
where $\ell_{vdW}$ is the natural-atomic length given by the van der Waals (vdW) part
of atomic potentials, introduces
some common behavior as was recently
justified~\cite{wang:2012_phys.rev.lett.,naidon:2012_arxiv:1208.3912[cond-mat.quant-gas]}. Motivated by
this fact we extract the values of $r_*$ and $r_+$ from the measured experiments. 
To this end, we can define $\kappa_*=(22.7)^{-1}1.56/|a^0_-|$ from the experimental value 
of $a^0_-$ to determine $a^1_{*,zr}$ and $a^1_{+,zr}$ and
then calculate the two lengths, which using Eq.~(\ref{eq:astar}) and
(\ref{eq:apiu}) can be expressed as
\begin{equation}
  \begin{aligned}
    r_* &= 1.032\,|a^0_-| - a^1_*\\
    r_+ &= 4.605\,|a^0_-| - a^1_+ \,.
  \end{aligned}
  \label{eq:calcoloRs}
\end{equation}
The results in units of $\ell_{vdW}$ are given in the last two rows of the
table. 
As general trends,  we can observe that the absolute value of the two lengths is of
order of 5$\;\ell_{vdW}$ (here potassium is the exception with a rather bigger
value of $r_+$), and that 
$|r_+|$ is slightly bigger than
$|r_*|$, but the values can be both positives and negatives. The results of 
the present work agree in order of magnitude with the experimental data, while the 
sign of of both $r_+$ and $r_*$ are probably related to the details of the 
specific Feshbach resonances. In the presented work, the resonance has been
explored by changing the strength of the potential, see Eq.~(\ref{eq:mtbp}),
which is a way to simulate a broad resonance but not a narrow one, and the
values of $r_+$ and $r_*$ we obtain are positives.

Though this analysis is not conclusive the results show that the
shifted formulae can be adapted to describe the experimental results after the
determination of $r_*$ and $r_+$.  Even if several efforts have been produced to
include finite-range corrections in the interpretation of experimental
data~\cite{frederico:1999_phys.rev.a,naidon:2012_phys.rev.a,ji:2010_epl,dincao:2009_j.phys.b,thogersen:2008_phys.rev.a,platter:2009_phys.rev.a},
our results suggest that new experimental as well as theoretical efforts are in
order to clarify this kind of corrections.

\begin{table}[h]
  \begin{tabular} { c|c c c c c }
    \hline
    \hline
         & $^7$Li\cite{dyke:2013_arxiv:1302.0281[cond-mat.quant-gas]}
         & $^7$Li\cite{machtey:2012_phys.rev.lett.} 
         & $^{39}$K\cite{zaccanti:2009_natphys,*roy:2013_phys.rev.lett.}
         &$^{133}$Cs\cite{ferlaino:2011_few-bodysyst.} 
         & present work    \cr
    \hline
    \hline
    $\ell_{vdW}(a_0)$ &  32.5 & 32.5 & 64.5 & 101.0 & 5.1 \cr
    \hline
    $a^0_-(a_0)$ & -241(8) & -280(12) & -690(40) &  -872(22)  & -48.1  \cr
    $a^1_-(a_0)$ &        &          &       &            & -975  \cr
  \hline
    $a^1_*(a_0)$ &  426(20) &  196(4) & 930(40) &  367(13)  & 16.0   \cr
  \hline
    $a^0_+(a_0)$ & 88(4) &           & 224(7)    & 210    &     \cr
    $a^1_+(a_0)$ & 1402(100)       & 1130(120) & 5650(900) &        & 135 \cr
  \hline
  $a^1_*/a^0_-$&  -1.77 &  -0.7     & -1.35      & -0.42  & -0.32 \cr
  $a^0_-/a^0_+$&  -2.7  &           & -3.08      & -4.2   & -7.2  \cr
  $a^0_-/a^1_+$&  -0.17 &  -0.25    & -0.12      &        & -0.35 \cr
  \hline
  $a^0_-/\ell_{vdW}$ & -7.4  &  -8.6     & -10.7     & -8.6   & -9.4  \cr
  $r_*/\ell_{vdW}$   & -5.5  &   2.9     &  -3.4       &  5.3   &  4.1  \cr
  $r_+/\ell_{vdW}$   & -9.0  &   4.9     &  -38.5    &  $<0$    &  6.2 \cr
  \hline
  \hline
  \end {tabular}
  \caption{Selected experimental results of the indicated experiments and
 selected ratios compared to the results of the present work
 (using the TBG+H3B potential).}
  \label{tab:expres}
\end{table}

\subsection{The nucleon-deuteron case}

To test the universal description of the recombination rate encoded in the 
threshold behavior of the ${\cal S}$-matrix we study a 
system which differs by order of magnitude on length and energy scales
with respect to  three-helium atoms. In particular, we shall consider the case of
neutron-deuteron ($n-d$) scattering.

An initial analysis for energies below the deuteron breakup threshold has been done
in Ref.~\cite{kievsky:2013_phys.rev.a} where it has been shown that the universal
formula of Eq.(\ref{eq:abcotdelta}) describes this process quantitatively
with $\kappa_*a+\Gamma_*\approx0.578$ and $a\approx 4.07\;$fm. These values have been estimated 
from the pole energy of $E_p=-160$ keV and the doublet $n-d$ scattering
length given in Ref.~\cite{chen:1989_phys.rev.c}. In this reference
calculations of $n-d$ scattering have been done using the $s$-wave spin
dependent potential of Malfliet and Tjon (MT) I-III model defined as
\begin{eqnarray}
V_t(r)&=&\frac{1}{r}\left( -626.885 e^{-1.55 r} +1438.72 e^{-3.11 r} \right)
\nonumber \\
V_s(r)&=&\frac{1}{r}\left( -513.968 e^{-1.55 r} +1438.72 e^{-3.11 r} \right)
\label{eq:mti-iii},
\end{eqnarray}
where $s$ and $t$ are the singlet and triplet spin states, $r$ is given in fm, 
and the potential in MeV. With $\hbar^2/m$=41.47 MeV fm$^2$ the triplet potential
leads to a binding energy for the deuteron $E_2=2.2307$ MeV. The predictions for
the singlet and triplet scattering lengths are $a_s=-23.583\;$fm and $a_t=5.513\;$fm
respectively, very close to the experimental results. The scattering lengths
are substantially bigger than the effective range of the system $r_s > 2\;$fm allowing 
an analysis in terms of the universal formulae. 

\begin{figure}
\begin{center}
  \includegraphics[width=\linewidth]{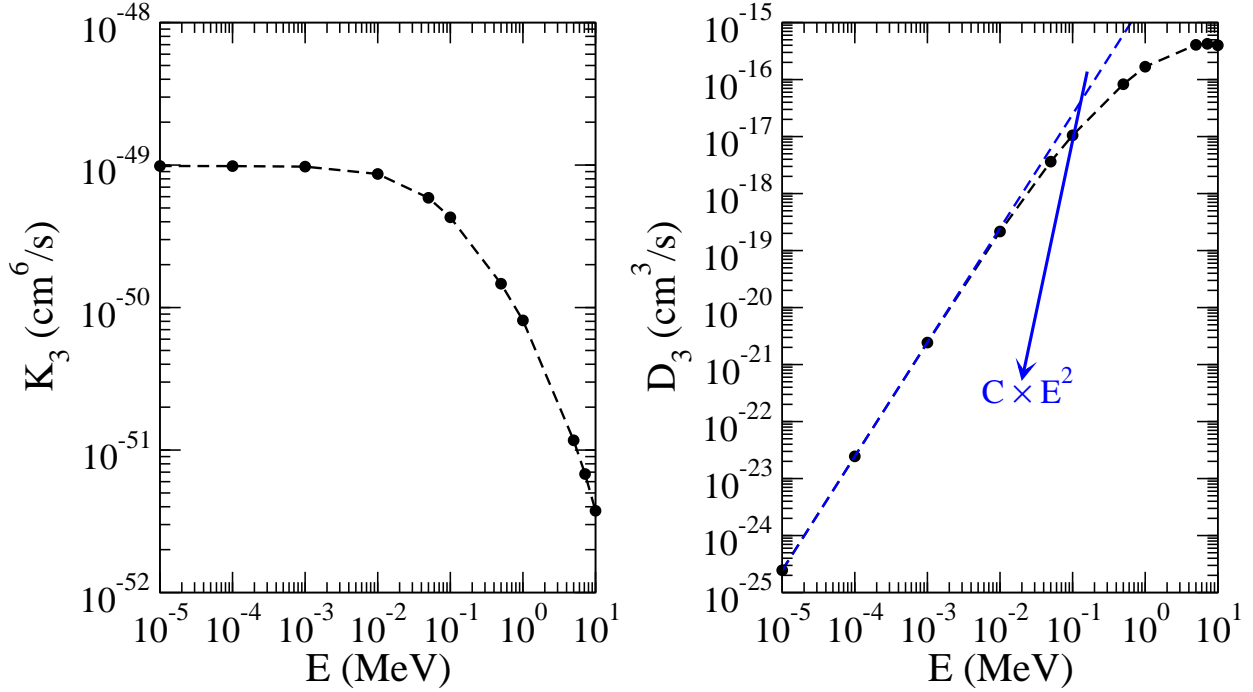}
  \caption{(Color online) The recombination rate $K_3$ and the dissociation rate
$D_3$ for  the $1/2^+$ state in the nucleon-deuteron system as a function of the energy.}
\label{fig:recomnd}
\end{center}
\end{figure}

Using the potentials above and the procedure described in
\cite{garrido:2012_phys.rev.a} we have calculated the corresponding
recombination and the dissociation rates, which are shown
in Fig.~\ref{fig:recomnd} as a function of the three-body energy $E$. The figure shows 
the $J=1/2^+$ state which is the only one having a $K_m=0$ contribution, being then the most
important one at low energies. When the threshold is approached ($E\rightarrow 0$) the rates
have the same behavior as in the atomic case, i.e., $K_3$ goes to a constant and $D_3$ 
goes to zero as $E^2$ (as shown by the dashed line in the right part of the
figure). At threshold we obtain $K_3=9.85\times 10^{-49}\;$ cm$^6$/s.

It is possible to analyze the behavior of $K_3$ at threshold in the context of the universal
function of Eq.(\ref{eq:abrecom}). However, it should be noticed that we can not simply replace
the value of the triplet scattering length, since both channels, the spin triplet
and the spin singlet, contribute to the process. To determine the value of the
scattering length to be used, let us have a look into the $J=1/2^+$
scattering wave function $\Psi_{nd}$, which is usually expanded in terms of a
complete set of adiabatic functions as~\cite{garrido:2012_phys.rev.a}
\begin{equation}
\Psi_{nd}(\rho,\Omega)= \frac{1}{\rho^{5/2}}\sum_{n=1}^{\infty}
f_n(\rho)\Phi_n(\rho,\Omega) ,
\end{equation}
where $\rho,\Omega$ are the hyperradial and hyperspherical coordinates and
the $\Phi_n(\rho,\Omega)$ functions are in fact the eigenfunctions of the
angular part of the Schr\"odinger (or Faddeev) equations. The angular
eigenfunctions are typically expanded in terms of the hyperspherical harmonics,
such that for our case with total angular momentum $L=0$, $\Phi_n(\rho,\Omega)$
is written as:
\begin{equation}
\Phi_n(\rho,\Omega)=  
 \sum_{K}[C_{K,0}(\rho)Y_K(\Omega)|s_x=0\rangle +
   C_{K,1}(\rho)Y_K(\Omega)|s_x=1\rangle ], 
\end{equation}
where $Y_K$ are HH functions, $C_{K,s_{x}}(\rho)$ are the coefficients in the
expansion, and we have made explicit the contributions from
the possible spin states $s_x=0,1$ of two nucleons, which after coupled to the spin
of the third nucleon give the total angular momentum of $1/2$.

The $n=1$ term describes asymptotically the elastic channel and $\Phi_1(\rho,\Omega)$ is proportional
to the deuteron wave function. The infinitely many remaining adiabatic channels describe the
breakup process \cite{garrido:2012_phys.rev.a}. Among them, the only one that asymptotically 
contains contributions
from hyperspherical harmonics with grand-angular momentum $K=0$ is the lowest adiabatic breakup 
channel (the $n=2$ term). This is the channel responsible for the behavior of the reaction
rates at threshold.  

\begin{figure}
\begin{center}
  \includegraphics[width=\linewidth]{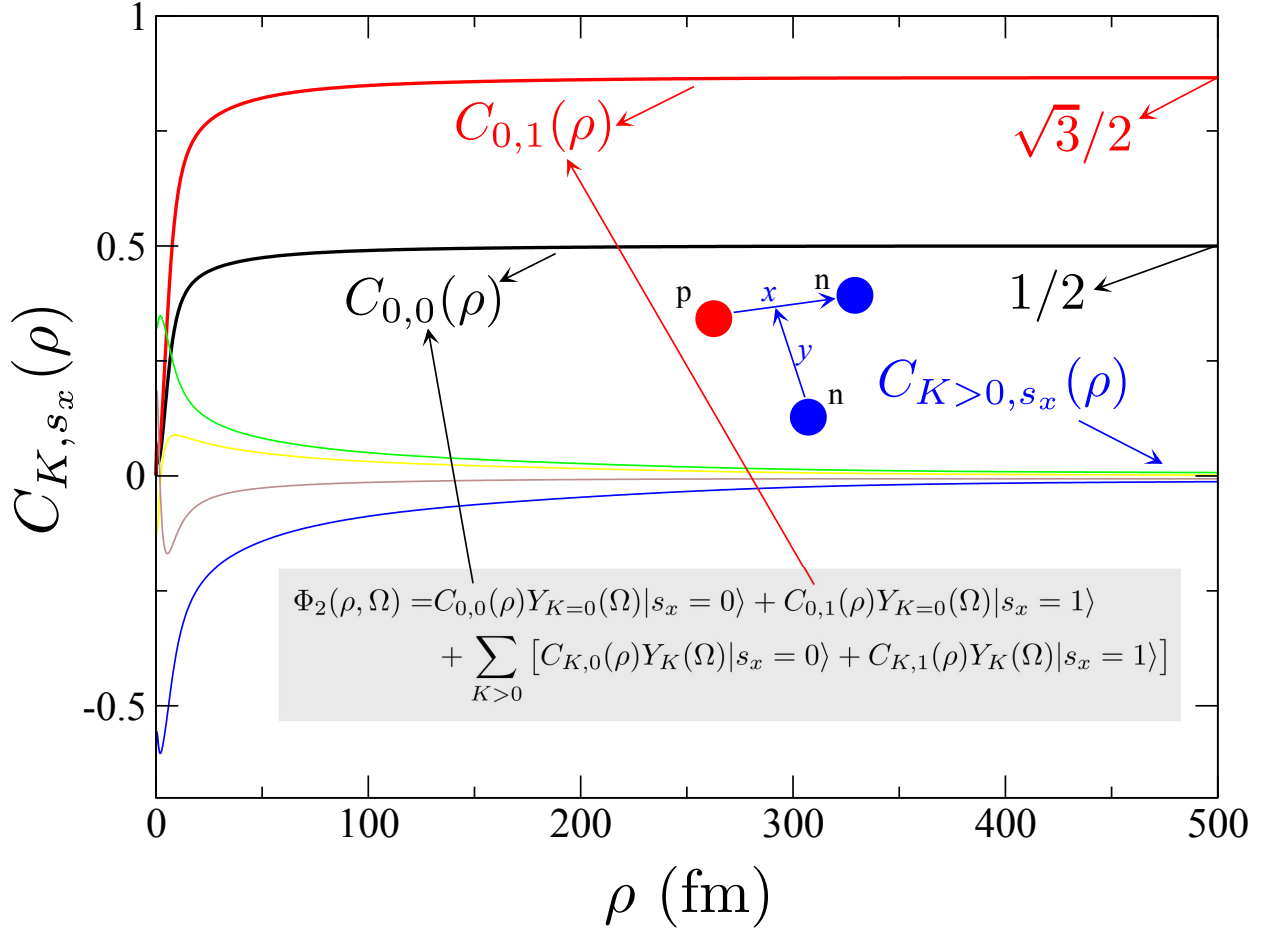}
  \caption{(Color online) The coefficients of the scattering wave function
as a function of the hyperradius $\rho$. The picture show the jacobi set of
coordinates used whereas the shadow box indicates the two coefficients
surviving at long distances.}
\label{fig:psind}
\end{center}
\end{figure}

In Fig.\ref{fig:psind} the coefficients of the $\Phi_2(\rho,\Omega)$ wave function are
shown as a function of the hyperradius. All of the coefficients with $K>0$ go to
zero as $\rho\rightarrow\infty$, but for $K=0$ the two coefficients 
$C_{0,0}(\rho)$ and $C_{0,1}(\rho)$ go to $1/2$ and $\sqrt{3}/2$, respectively. In the
figure the Jacobi coordinates have been depicted indicating by $x$ the $n-p$
relative distance and by $y$ the relative distance of the third nucleon
to the $n-p$ center of mass, so $s_x=0,1$ is the $n-p$ spin. The shadow
box shows explicitly the two surviving terms when $\rho$ increases.

This analysis implies that $K_3$ receives contribution from the singlet
and triplet channel properly weighted. Therefore in the universal formula
the scattering length $a$ has to be considered as an effective scattering
length $a=12.68$ fm, obtained from the average $a^2= (3a_t^2+a_s^2)/4$.
Moreover, the value of $a_B$ results $12.37\;$fm if in the average
$a_t$ is replaced by $\hbar/\sqrt{mE_2}$. With the values for $a$, $a_B$ and 
the product $\kappa_*a+\Gamma_+\approx1.8 $ (obtained from the $n-d$ value 
of 0.578 multiplied to the ratio of $a/a_t$), we
obtain from Eq.(\ref{eq:abrecom}) the recombination rate value of 
$K_3=9.5\times 10^{-49}\;$ cm$^6$/s, very close to the computed
value. This qualitative analysis can be used as an starting point in
the spin analysis of the universal aspects of the recombination rate
in light nuclear systems.

\section{conclusions}

In the present paper we have investigated atom-dimer scattering above 
the breakup threshold using a set of gaussian potentials constructed
to reproduce the low-energy atom-atom system. We have introduced a
three-body force in order to reproduce the trimer binding energies
given by the LM2M2 potential multiplied by a constant $\lambda$ varied
from $\lambda\approx 0.974$ to $\lambda\approx1.1$ in order to cover
a large part of the $(a^{-1},\kappa)$ plane. We have studied the total elastic and
the total breakup cross sections, as well as the recombination and dissociation 
rates, $K_3$ and $D_3$, which are directly related to the latter. We have payed
special attention to the behavior of the system close to threshold, and
we have investigated the universal behavior of $K_3$ for different values
of the two-body scattering length $a$. Our aim here was to analyze
the differences between the universal zero-range theory which, for $K_3$
is given by Eq.(\ref{eq:recomb}), and more realistic finite range calculations.
This study started already in
Ref.~\cite{kievsky:2013_phys.rev.a}, where this kind
of differences have been studied for the first excited level of the
three-atom system and in the effective range function. In that reference
it was found that finite-range results organize in a curve shifted with
respect to the zero-range theory. Here we have extended the analysis to
the ground state energy and to the recombination rate. Our main results
are given in Fig.~\ref{fig:recomth} where we can observe that the
computed values for $K_3$ lie on a shifted curve in the variable 
$\kappa_*a$. Additional range corrections come from the replacement
of $a$ by $a_B$ in the $\hbar a^4/m$ factor. This type of correction
has been already introduced in the study of $a_{AD}$ and $k\cot\delta$
as well as in the binding energies $E_3^0$ and $E_3^1$ which have 
been divided by $E_2$ instead of $\hbar^2/ma^2$, as indicated by the
zero-range theory.

The fact that the finite-range results lie on a shifted curve
in the variable $\kappa_*a$ for the binding energies and the
low-energy scattering quantities strongly support this 
particular type of correction. Following this observation, we have proposed 
Eqs.(\ref{eq:ab_AD}), (\ref{eq:abcotdelta}), and (\ref{eq:abrecom}) as
modifications to the zero-range universal formulae where we
have introduced the shifts $\Gamma_*,\Gamma_e$ and $\Gamma_+$. 
From the calculations we get that the shift of the first excited 
$\Gamma_1\approx\Gamma_*\approx\Gamma_e$ whereas $\Gamma_+$ is slightly
larger. We have argued that in the range of $a$ studied the three-boson
system has only one excited state making the shift of the same order.
In the case of the recombination rate one more channel is open and,
in terms of adiabatic potentials, at least two channels have to be considered
(see for example Section 6 of Ref.~\cite{braaten:2006_physicsreports}).
As a consequence a larger amount of potential energy appears increasing
the shift. This intuitive arguments can be stated formally; work along
this line is in progress.

Most of the work presented here has been focussed in the necessity of
adapting the universal formulae obtained in a zero-range theory to the 
physical case of finite-range interactions. Accordingly, the modifications
introduced in the three-helium system model similar modifications to be
considered in the analysis of the experiments on Efimov resonance in ultracold
gases. The analysis of the experimental results given in Section III.C was
directed to see if there is experimental evidence of the lengths
$r_*$ and $r_+$ introduced in the parametrization of the shift.
We have observed some common behavior of the different species when
the lengths are given in units of $\ell_{vdW}$. This fact encourages
this kind of analysis and suggests that more theoretical as well 
experimental efforts can be done in this direction.

In the last part of this study we have focused on the study of
nucleon-deuteron scattering closely above the breakup threshold. 
Using the spin dependent MT I-III potential we have
shown that the $K_3$ and $D_3$ rates have a behavior close to
threshold similar to the one observed in the three-atom systems.
We have applied the universal formula to this system using
a spin average scattering length and we have seen that 
it was possible to describe the computed value of $K_3$. This qualitative
analysis can be considered as a first step in the study of universal
aspects of nucleon-deuteron scattering considering the full spin
dependence of the system. 
\acknowledgments E.G. acknowledges financial support provided by
DGI of MINECO (Spain) under contract No. FIS2011-23565. 

\end{document}